\documentclass[aps,prb,twocolumn,superscriptaddress,showpacs]{revtex4}

\usepackage{graphicx}
\usepackage{amssymb}

\begin{document}

\title{Band-filling effects on the Kondo-lattice properties}

\author{B. Coqblin}
\affiliation{Laboratoire de Physique des Solides, Universit\'{e}
Paris-Sud, B\^atiment 510, 91405 Orsay, France}
\author{C. Lacroix}
\affiliation{Laboratoire L. N\'{e}el, CNRS, BP 166, 38042 Grenoble
Cedex 09, France}
\author{M. A. Gusm\~ao}
\author{J. R. Iglesias}
\affiliation{Instituto de F\'{\i}sica, Universidade Federal do Rio
Grande do Sul, C.P. 15051, 91501-970 Porto Alegre, Brazil}

\date{\today}

\begin{abstract}
We present theoretical results for a Kondo-lattice model with spin-$1/2$
localized moments, including both the intrasite Kondo coupling and an
intersite antiferromagnetic exchange interaction, treated within an extended
mean-field approximation. We describe here the case of a non-integer
conduction-band filling for which an ``exhaustion'' problem arises when the
number of conduction electrons is not large enough to screen all the lattice
spins.  This is best seen in the computed magnetic susceptibility. The Kondo
temperature so obtained is different from the single-impurity one, and
increases for small values of the intersite interaction, but the
Kondo-effect disappears abruptly for low band filling and/or strong
intersite coupling; a phase diagram is presented as a function of both
parameters. A discussion of experimental results on cerium Kondo compounds
is also given.
\end{abstract}


\maketitle

\section{\label{sec:intro}INTRODUCTION}

The properties of many cerium or ytterbium compounds are well accounted for
by the Kondo-lattice model, where a strong competition exists between
the Kondo effect and magnetic ordering arising from the RKKY
(Ruderman-Kittel-Kasuya-Yosida) interaction between rare-earth atoms at
different lattice sites. This situation is well described by the Doniach
diagram,\cite{Doniach} which gives the variation of the N\'eel temperature
and of the Kondo temperature with increasing antiferromagnetic intrasite
exchange interaction $J_K$ between localized spins and conduction-electron
spins. If one considers the exchange Hamiltonian between localized ($\bf S$)
and conduction-electron ($\bf s$) spins, given by
\begin{equation} \label{eq:Kexch}
  H =  J_K \, {\bf s} \cdot {\bf S} \,,
\end{equation}
usual theories of the one-impurity Kondo effect and of the RKKY interaction
yield a Kondo temperature $T_{\rm K0}$ that is proportional to $\exp (-
1/{\rho}J_K)$, and an ordering temperature (N\'eel or, in some cases, Curie)
$T_{\rm N0}$, proportional to ${\rho}J_K^2$, ${\rho}$ being the density of
states for the conduction band at the Fermi energy. Thus, for small
${\rho}J_K$ values, $T_{\rm N0}$ is larger than $T_{\rm K0}$ and the system
tends to order magnetically, with often a reduction of the magnetic moment
due to the Kondo effect. On the contrary, for large ${\rho}J_K$, $T_{\rm
K0}$ is larger than $T_{\rm N0}$ and the system tends to become non
magnetic. The actual ordering temperature $T_{N}$, therefore, increases
initially with increasing ${\rho}J_K$, then passes through a maximum and
tends to zero at a critical value ${\rho}J_K^c$ corresponding to a ``quantum
critical point'' (QCP) in the Doniach diagram. Such a behavior of $T_{N}$
has been experimentally observed with increasing pressure in many cerium
compounds, such as CeAl$_{2}$ (Ref.~\onlinecite{Barbara}) or
CeRh$_{2}$Si$_{2}$.\cite{Graf} We also know that the N\'eel temperature
starts from zero at a given pressure, and increases rapidly with pressure in
YbCu$_{2}$Si$_{2}$ (Ref.\ \onlinecite{Alami}) or in related ytterbium
compounds, which can be considered as another test of the Doniach
diagram. The one-impurity model predicts an exponential increase of the
Kondo temperature with ${\rho}J_K$. This means that the Kondo temperature
should increase with increasing pressure in cerium compounds and with
decreasing pressure in ytterbium compounds, in good agreement with many
observations. However, deviations seem to occur in some cerium compounds,
such as CeRh$_{2}$Si$_{2}$ (Ref.~\onlinecite{Graf}) or
CeRu$_{2}$Ge$_{2}$,\cite{Sullow,Wilhelm} where the actual Kondo temperature
observed in a lattice can be significantly different from the one derived
for the single-impurity case. Thus, in order to account for such an effect,
we have treated in a previous paper\cite{Iglesias} the Kondo-lattice model
with both intrasite Kondo exchange and intersite antiferromagnetic exchange
interactions, for a half-filled conduction band (corresponding to a number
of conduction electrons $n=1$). We employed a mean-field approximation with
two correlators, $\lambda^{}_{i\sigma}$, describing the intrasite Kondo
correlation, and $\Gamma^{}_{ij\sigma}$, representing an intersite
correlation between two neighboring moments. We have shown that the
enhancement of the intersite exchange interaction tends to decrease the
Kondo temperature $T_{K}$ for the lattice with respect to the one-impurity
Kondo temperature $T_{\rm K0}$, and to suppress the Kondo effect for large
values of the intersite exchange interaction parameter.\cite{Iglesias} So,
this model can account for the pressure dependence of $T_{K}$ observed in
CeRh$_{2}$Si$_{2}$,\cite{Graf} CeRu$_{2}$Ge$_{2}$,\cite{Sullow,Wilhelm} or
more recently\cite{Umeo} Ce$_{2}$Rh$_{3}$Ge$_{5}$.

The above mentioned model also yields a correlation temperature $T_{\rm
cor}$, below which short-range magnetic correlations between neighboring
cerium atoms occur. For sufficiently large values of the intersite exchange,
$T_{\rm cor}$ is higher than the Kondo temperature. This result is in
agreement with the experimental observation by neutron diffraction
experiments of such short-range magnetic correlations in single crystals of
CeCu$_{6}$,\cite{Rossat} CeInCu$_{2}$,\cite{Pierre}
CeRu$_{2}$Si$_{2}$,\cite{Flouq,Rossat,Regnault} or
Ce$_{1-x}$La$_{x}$Ru$_{2}$Si$_{2}$ (Refs.\
\onlinecite{Flouq,Regnault,Mignot,Regnault2}) at low temperatures.  The
experimentally observed temperature $T_{\rm cor}$ is clearly larger than the
Kondo temperature $T_{K}$: $T_{\rm cor} {\sim}$ 60--70 K and $T_{K} {\sim}$
20 K in CeRu$_{2}$Si$_{2}$;\cite{Rossat,Flouq,Regnault} $T_{\rm cor} {\sim}$
10 K and $T_{K}{\sim}$ 5 K in CeCu$_{6}$.\cite{Rossat}

Most Kondo-lattice models have been studied for the case of a half-filled
conduction band, corresponding to a number of conduction electrons, $n$,
equal to the number $n_{f} = 1$ of $f$ electrons. However, when $n < 1$, an
``exhaustion'' problem arises, which means that there are not enough
conduction electrons to screen all the localized spins and, as a
consequence, the Kondo temperature decreases.\cite{Lacroix-79,Nozieres} The
stability of the Kondo effect when $n$ decreases has been recently
studied,\cite{Alaor} and also the temperature dependence of the magnetic
susceptibility has been computed for different values of $n$,\cite{Coq-2000}
with the Hamiltonian including an intersite exchange interaction $J_{H}$
between the localized spins, treated as independent of $J_K$. The results
showed that a reduction of $n$ and/or an enhancement of $J_{H}$ tend to
suppress the Kondo effect. In particular, the analytical calculation at
$T=0$ gave $\lambda _{}^{2}$ proportional to $n$ in the case of
$J_{H}=0$.\cite{Coq-2001}

Early studies of the Kondo-lattice model have considered only
the intrasite Kondo interaction.\cite{Jullien,Continentino,Tsunetsugu} In
particular, Continentino et al.,\cite{Continentino} using scaling theory,
have found a coherence temperature increasing above the QCP. Recently,
Burdin et al.,\cite{Burdin,Burdin2} using the slave-boson method in the
exhaustion limit, have obtained two different energy scales: the
single-impurity Kondo temperature $T_{K}$, which corresponds to the onset of
local singlet formation, and the zero-temperature energy gain, $T^{*}$,
related to the coherent Kondo effect. Thus, the low-temperature energy scale
is not equal to the high-temperature one, although there is only one
mean-field parameter in the theory.\cite{Burdin,Burdin2} It has been shown
that the ratio $T^{*}/T_{K}$ is much smaller than 1, and depends only on the
band filling factor. In the case of a rectangular density of states for the
conduction band, $T_{K}$ vanishes as $n^{1/2}$ and the coherence temperature
as $n$ in the limit $n \to 0$. However, both $T_{K}$ and $T^{*}$ where found
to exhibit the same exponential behavior $\exp{(-2D/J_{K})}$, in contrast
with the formula established by Nozi\`eres,\cite{Nozieres} who found $T^{*}
\sim T_{K}^{2}/D$. The effect of a small number of conduction electrons has
been also studied within both the Kondo-lattice and the Anderson-lattice
models.\cite{Castro,Ben-Hur,Tahvildar,Arko}

The purpose of the present paper is to further develop the study of
conduction-band filling effects within the Kondo-lattice model.  Besides
previous work cited above,\cite{Iglesias,Coq-2000,Coq-2001} a first brief
account has been already given,\cite{Lacroix-2001} but we will present here
both numerical results and new analytical expressions for some specific
limits, as well as a detailed discussion of band-filling effects.

\section{\label{sec:revisit}The Theoretical Model}

As it was mentioned above, the Kondo-lattice model is often
studied with only the intrasite Kondo interaction. In principle,
both the RKKY magnetic interaction and the Kondo effect are
consequences of the intrasite exchange term. However, the RKKY
interaction is perturbative in $J_K$ while the Kondo effect is
not.\cite{Castro} When dealing with approximation schemes, it is
hard to obtain both effects starting with the intrasite
interaction alone. For this reason, an explicit intersite exchange
interaction is usually included in the model
Hamiltonian.\cite{Iglesias,Castro,Coleman}

The proposed Hamiltonian of the system is, therefore,
\begin{equation} \label{eq:model}
H = \sum_{{\bf k}\sigma} \varepsilon_{\bf k}^{} n_{{\bf k}\sigma}^{c} + J_K
\sum_i {\bf s}_i \cdot {\bf S}_i + J_H \sum_{\langle ij \rangle} {\bf S}_i
\cdot {\bf S}_j \;,
\end{equation}
where $\varepsilon_{\bf k}$ is the energy of the conduction band, $J_{K}$ is
the Kondo coupling between a localized spin ${\bf S}_i$ and the spin ${\bf
s}_i$ of a conduction electron at the same site, and $J_{H}$ is the
interaction between nearest-neighboring localized spins. Assuming spin-1/2
localized moments, we represented them by a zero-width $f$ band with one
electron per site, while the conduction band has width $2D$ and a constant
density of states. We choose $J_{K}$ and $J_{H}$ to be positive, implying
that both local and intersite interactions are antiferromagnetic, as is the
case in most cerium compounds.

We now write the spin operators in fermionic representation:
\begin{eqnarray} \label{eq:spins}
s_i^{z} = \frac{1}{2} \left( n^c_{i\uparrow} - n^c_{i\downarrow}
\right)\:, &&\qquad S_i^{z} = \frac{1}{2} \left( n^f_{i\uparrow} -
n^f_{i\downarrow} \right)\:, \nonumber \\ s_i^{+} =
c_{i\uparrow}^\dagger c_{i\downarrow} \:, &&\qquad S_i^{+} =
f_{i\uparrow}^\dagger f_{i\downarrow} \:, \\ s_i^{-} =
c_{i\downarrow}^\dagger c_{i\uparrow}  \:, &&\qquad S_i^{-} =
f_{i\downarrow}^\dagger f_{i\uparrow} \:, \nonumber
\end{eqnarray}
remembering that we have a constraint of single-occupancy of the
$f$ level at all sites, $n_i^f=1$.

In order to discuss the Kondo effect and magnetic correlations we
define the operators
\begin{equation} \label{eq:lamgam}
\lambda^{}_{i\sigma}  \equiv
c_{i\sigma}^\dagger f_{i\sigma}^{} \:, \quad
\Gamma^{}_{ij\sigma}  \equiv
f_{i\sigma}^\dagger f_{j\sigma}^{} \: ,
\end{equation}
where $\lambda^{}_{i\sigma}$ describes the intrasite Kondo
correlation, and $\Gamma^{}_{ij\sigma}$ represents an intersite
correlation between two neighboring atoms. This is so because the
on-site spin-spin correlation $\langle {\bf s}_i \cdot {\bf S}_i
\rangle$ can be written in terms of the average $\langle
\lambda^{}_{i\sigma} \lambda^{}_{i,-\sigma}\rangle$, and similarly
for the intersite correlation using $\Gamma^{}_{ij\sigma}$. With
this notation we perform an extended mean-field approximation,
introduced by Coleman and Andrei,\cite{Coleman} and presented in
full detail in Ref.\ \onlinecite{Iglesias}. Considering
translational invariance, and taking into account that there is no
breakdown of spin symmetry, i.e., no magnetic states, we can write
$\lambda = \langle \lambda_{i\sigma} \rangle$ for all sites, and
$\Gamma = \langle \Gamma_{ij\sigma} \rangle$ for
nearest-neighboring sites and zero otherwise. In this way we
obtain a mean-field Hamiltonian that takes the form of a
hybridized two-band system:
\begin{eqnarray} \label{eq:HMF}
H_{}^{\rm MF} &=& \sum_{{\bf k}\sigma} \varepsilon_{\bf
k} n^c_{{\bf k}\sigma} + E_0 \sum_i \left( \sum_\sigma n^f_{i\sigma} - 1
\right) \nonumber
\\ && \mbox{} - J_K \lambda \sum_{i\sigma} \left(
c_{i\sigma}^\dagger f_{i\sigma}^{} + f_{i\sigma}^\dagger
c_{i\sigma}^{} \right) + \bar{E}_K \nonumber \\ &&
\mbox{} - J_H \Gamma \sum_{\langle ij\rangle \sigma}
\left( f_{i\sigma}^\dagger f_{j\sigma}^{} + f_{j\sigma}^\dagger
f_{i\sigma}^{} \right) + \bar{E}_H \:,
\end{eqnarray}
with
\begin{equation} \label{eq:Ebarsfinal}
\bar{E}_K =  2 N J_K \lambda^2 \:, \quad
\bar{E}_H =  z N J_H \Gamma^2 \:,
\end{equation}
$N$ being the total number of lattice sites. We have introduced a term
depending on the Lagrange multiplier $E_0$ in order to impose the constraint
$\sum_i (n_i^f-1)=0$. This is a weak form of the original constraint
$n_i^f=1$.

After performing this approximation, one deals with a one-electron
Hamiltonian representing two hybridized bands: the conduction
band of width $2D$ and the $f$ band of effective band width $2BD$, with
$B$ given by
\begin{equation} \label{eq:zJH}
B= -z J_H \Gamma/D,
\end{equation}
$z$ being the number of nearest neighbors of a site, while the magnitude of
the hybridization gap is directly related to $\lambda _{}^{2}$.  This
quantity is also a measure of the Kondo effect, as the Kondo correlation
function $\langle {\bf s}_i \cdot {\bf S}_i \rangle$ is proportional to
${\lambda}^{2}$.

The Hamiltonian (\ref{eq:HMF}), leaving aside the constant terms, is easily
diagonalized, and the resulting energies of the two new hybrid bands are
\begin{eqnarray} \label{eq:eigen}
E_{\bf k}^{\pm} &=& \frac{1}{2} \bigg[ \,\varepsilon_{\bf k}^{} (1+B)
+ E_0 \nonumber \\ && \mbox{~} \pm \sqrt{\left[\,
\varepsilon_{\bf k}^{} (1-B) - E_0 \right]^2 + 4 J_K^2
\lambda^2} \; \bigg] \:
\end{eqnarray}

The mean-field parameters $\lambda$ and $\Gamma$ are obtained by
self-consistently solving Eqs.~(\ref{eq:lamgam}) or, equivalently, by
minimizing the total internal energy
\begin{equation} \label{eq:energ}
E = 2\sum_{{\bf k}, \alpha=\pm} E_{\bf k}^\alpha f(E_{\bf
k}^\alpha)  + \bar E_K + \bar E_H - E_0 N
\end{equation}
at zero temperature, or the total free energy $F$ at finite
temperatures.\cite{Alaor} The summation in Eq.~(\ref{eq:energ}) is made over
all {\bf k} states in the first Brillouin zone, and $f(E_{\bf k})$ is the
Fermi-Dirac function. As usual,\cite{Iglesias} the reference energy $E_{0}$
of the $f$ band and the chemical potential $\mu$ have to be determined
self-consistently in order to keep the average numbers of $f$ and conduction
electrons respectively equal to 1 and $n$.

We have previously treated the half-filled case, corresponding to a number
of conduction electrons $n = 1$,\cite{Iglesias} and here we consider the
general case of a non-integer number of conduction electrons. For the
half-filled case,\cite{Iglesias} we have shown that the Kondo temperature
$T_{K}$ for the lattice can be much smaller than the Kondo temperature
$T_{K0}$ for the single impurity, as observed in some cerium compounds. We
will present in the next sections the theoretical results for band-filling
effects, and discuss later on the comparison with experiment.

\bigskip
\section{\label{sec:filling}Band-filling effects at $T=0$}

The two hybridized bands $E_{\bf k}^{\pm}$ given by Eq.\ (\ref{eq:eigen})
exhibit a structure that depends on the two factors $A\equiv J_K^2
\lambda{^2}$ and $B$, defined by (\ref{eq:zJH}), and especially on the sign
of $B$. In the case of $n = 1$, previously considered,\cite{Iglesias} for
small $|B|$ values such that $|B|D^2<A$ there is a gap, the lower band is
completely full and the upper band empty at $T = 0$. For $n < 1$, the Fermi
level cuts the lower band, and the upper band is still empty at $T = 0$,
which corresponds, therefore, to a real metallic situation.

The shape of the lower band $E_{\bf k}^{-}$ and the solution of
the case $ n < 1$ depend critically and self-consistently on the
values of the different parameters $J_{K}$, $J_{H}$ and $n$ or
equivalently $A$, $B$ and $n$. If $B > 0$, i.e., $ \Gamma < 0$
(since $J_{H}$ is positive here), both the ``$f$ band'' and the
hybridized lower band $E_{\bf k}^{-}$ are continuously increasing
with $\varepsilon_{\bf k}$. The same structure exists for $B < 0$
and small $J_{H}$ values, while $E_{\bf k}^{-}$ presents a maximum
for $B < 0$ and larger $J_{H}$ values.

\begin{figure}
\includegraphics[width=8.5cm,clip]{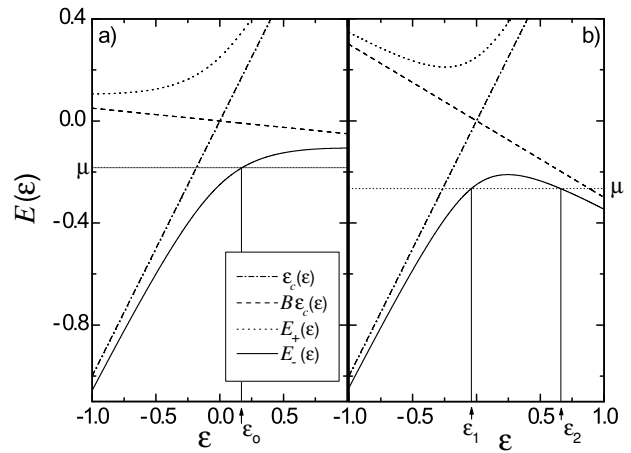}
 \caption{\label{fig:enercurv}Schematic plot of the non-hybridized conduction
 and 4$f$ bands, and of the two hybridized bands $E_{\pm}({\bf k})$. For a
 given $n<1$, the chemical potential is different in panels a) and b), which
 correspond respectively to the two cases studied: small $J_H$ (a single
 solution of $E_{-}(\varepsilon)=\mu)$, and large $J_H$ (two solutions). All
 energies appear in units of $D$.}
 \end{figure}

Thus, the shape of $E_{\bf k}^{-}$ can change under a variation of the
parameters, and this peculiar situation makes the problem difficult to
solve. The two characteristic cases are presented schematically in
Fig.~\ref{fig:enercurv}, where we have used the fact that the energies
$E_{\bf k}^{\pm}$ depend on $\bf k$ only through the bare conduction-band
energies $\varepsilon ({\bf k})$, making the substitution $E_{\bf k}^{\pm}
\to E_{\pm} (\varepsilon)$, with $\varepsilon$ defined in the interval
[-1,1] in units of the half bandwidth $D$. As can be seen in
Fig.~\ref{fig:enercurv}, the curve $E_{-} (\varepsilon)$ may or may not have
a maximum, according to the different cases studied. Moreover, when $E_{-}
(\varepsilon)$ presents a maximum, the Fermi energy can cut the lower band
in two points or in only one point within the first Brillouin zone. In fact,
we have to distinguish two cases, depending on whether there are one or two
intersection points (which implies automatically in the second case that the
curve $E_{-} (\varepsilon)$ has a maximum). We will compute the total energy
in both situations.

For $T=0$, we firstly derive analytical expressions for both cases, and then
we present numerical solutions for different sets of parameters that affect
the shape of $E_{-} (\varepsilon)$.  In the first case, shown in
Fig.~\ref{fig:enercurv} a), the energy per lattice site is written as
\begin{eqnarray} \label{eq:nomax}
\mathcal{E} & = & \frac{1}{2D}  \nonumber \int_{-D}^{\varepsilon
_{0}} d\varepsilon \biggl[ E_{0} +(1+B)\varepsilon  \nonumber \\
&& \mbox{} - \left. \sqrt{\left[E_{0} -(1-B)\varepsilon
\right] ^{2} +4J_{K}^{2} \lambda ^{2}} \right] \\
&& \mbox{}  + zJ_{H} \Gamma^{2} + 2J_{K} \lambda ^{2} - E_{0} \;. \nonumber
\end{eqnarray}
The upper limit $\varepsilon_{0}$ is the value of $\varepsilon_{\bf k}$ that
corresponds to $E_{\bf k}^{-}= \mu$ in Eq.~(\ref{eq:eigen}), as shown in
Fig.~\ref{fig:enercurv}. Writing the two self-consistent equations that
yield $n_{f}=1$ and the number of conduction electrons $n$, we obtain
$\varepsilon_{0} = Dn$, and
\begin{equation} \label{eq:Eo}
E_{0} =-\frac{1}{2}(1-n) \left[(1-B)\;D + \Delta \right] \;,
\end{equation}
with
\begin{equation} \label{eq:Delta}
\Delta =\sqrt{D^{2} (1-B)^{2} +\frac{4J_{K}^{2} \lambda^{2}}{n}} \;.
\end{equation}

Finally, the internal energy per site at $T=0$ [Eq.~(\ref{eq:nomax})] is
given by
\begin{eqnarray} \label{eq:enm}
\mathcal{E} & = & zJ_{H} \Gamma ^{2} + 2J_{K} \lambda^{2} - E_0 + \frac{D}{2} (n^{2}
 -1)-\frac{\Delta }{2} \nonumber\\ & & +\frac{J_{K}^{2} \lambda ^{2}
 }{D(1-B)} \ln \left\{ \frac{\Delta -D(1-B)}{\Delta +D(1-B)}\right\} \;.
\end{eqnarray}

The derivatives of the total energy with respect to $\lambda$ and $\Gamma$
yield two self-consistent equations, whose solutions give the final values
of $\lambda$ and $\Gamma$ corresponding to the minimum energy. For
small values of $J_{H}$, Eq.~(\ref{eq:enm}) allows us to write
\begin{eqnarray}
\lambda _{}^{2}  &=& \frac{n D_{}^{2} (1-B)_{}^{2}}{J_{K}^{2}} \,
\frac{u}{(1-u)_{}^{2}} \;,\label{eq:lamn}\\
\Gamma  &=& \frac{n\Delta}{4D(1-B)} - \frac{1-n}{4} - \frac{J_{K}
\lambda _{}^{2} }{D(1-B)} \;, \label{eq:gamn}
\end{eqnarray}
with $ u \equiv \exp\left[- 2D(1-B)/J_{K} \right]$.

The case with a maximum in $E_{\bf k}^{-}$ is more complicated. As shown in
Fig.~\ref{fig:enercurv}, for the case $B < 0$ with large values of $J_{H}$,
the Fermi level $\mu$ cuts the band $E_{\bf k}^{-}$ in two energies; then,
the two corresponding energies $\varepsilon_{1}$ and $\varepsilon_{2}$ are given
by the values of $\varepsilon({\bf k})$ which are roots of the equation $E_{\bf
k}^{-} = \mu$ with $E_{\bf k}^{-}$ given by eq.(\ref{eq:eigen}). So, the
expression of the energy is given by Eq.~(\ref{eq:nomax}) by changing the
integration limits so that the integral is now performed from $-D$ to
$\varepsilon_{1}$ and from $\varepsilon_{2}$ to $D$.

Now, the two equations derived from the constraint $n_{f} = 1$ and
the number of conduction electrons equal to $n$ are
\begin{equation} \label{eq:epsilon2}
\varepsilon_{2} - \varepsilon_{1} = D (1 - n) \;,
\end{equation}
\begin{equation} \label{eq:Eomax}
E_{0} = - (1 - n)\sqrt{D^{2} +\frac{4J_{K}^{2} \;\lambda^{2}
}{(1-B)^{2} -(1-n)^{2} } } \;.
\end{equation}

We can derive analytical expressions for $\varepsilon_{1}$ and
$\varepsilon_{2}$, and finally for the total energy per site $\mathcal{E}$
for $B < 0$. A quite long calculation yields
\begin{widetext}
\begin{eqnarray} \label{eq:Emax}
\mathcal{E} & = & -\frac{R}{2(1-B)} \left[ (1-B)^{2} -2B(1-n)+(1-n^{2}
)\right] -\frac{J_{K}^{2} \lambda ^{2} }{(1-B)D} \ln\left\{
\frac{P-D(1-n)\sqrt{\left| B\right| } }{P+D(1-n)\sqrt{\left| B\right| } }
\cdot \frac{R+D}{R-D} \right\}\nonumber \\ & & \mbox{} + \frac{1-n}{2(1-B)}P
\sqrt{\left| B \right|} + \frac{D^{2} B^{2} }{zJ_{H} } + 2J_{K} \lambda ^{2}
- E_{0}\;,
 \end{eqnarray}
\end{widetext}
where
\begin{equation} \label{eq:R}
R=\sqrt{D^{2} +\frac{4J_{K}^{2} \;\lambda ^{2} }{(1-B)^{2}
-(1-n)^{2}}}
\end{equation}
and
\begin{equation} \label{eq:P}
P=\sqrt{4J_{K}^{2} \,\lambda ^{2} -D^{2} B(1-n)^{2}} \;.
\end{equation}
Minimizing the energy (\ref{eq:Emax}) with respect to
${\lambda}$ we obtain
\begin{equation} \label{eq:lambda}
\lambda _{}^{2} = p/2a \;,
\end{equation}
with
\begin{equation} \label{eq:p}
p = 2J_{K} - \frac{J_{K}^{2} }{D(1-B)} \ln \left[ \frac{(1-B)_{}^{2}
-(1-n)_{}^{2} }{-B(1-n)_{}^{2} }\right] \;,
\end{equation}
and
\begin{equation} \label{eq:a}
a =  \frac{J_{K}^{2}}{4D(1-B)} \left[
\frac{1}{B(1-n)_{}^{2}} + \frac{1}{(1-B)_{}^{2} - (1-n)_{}^{2}}
\right] \;.
\end{equation}
For the present case, ${\lambda}$ vanishes at a critical band filling given
by
\begin{equation} \label{eq:nc}
n_{c} = 1-(1-B) \sqrt{\frac{u}{u-B}} \;.
\end{equation}

We have finally performed a numerical calculation by considering the general
form of the energy given by either Eq.~(\ref{eq:nomax}) or
Eq.~(\ref{eq:Emax}) in the different cases studied, and looking for the
minimum energy of the system. In fact, for each value of the different
parameters, we compare the two solutions and we retain the lowest-energy
one, thus determining the corresponding $\lambda$ and $\Gamma$.

\begin{figure}
\includegraphics[width=8.5cm,clip]{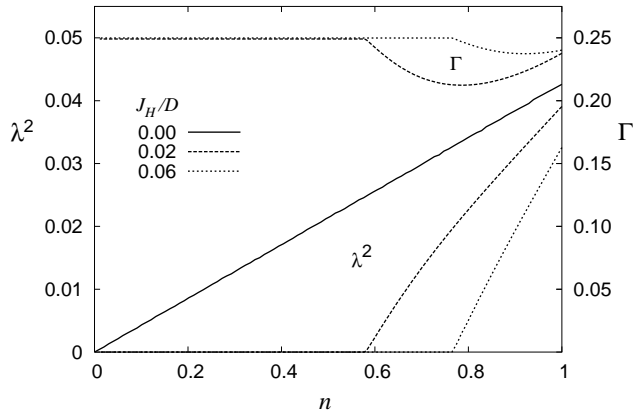}
\caption{\label{fig:kondoxn}Variation of $\lambda^{2}$ and ${\Gamma}$ with
  the band filling $n$ for $J_{K}/D = 0.4$ and different values of $J_{H}$.}
\end{figure}

Figure \ref{fig:kondoxn} shows $\lambda^{2}$ and $\Gamma$ versus $n$ for
different values of $J_{H}$, and $J_{K}/D=0.4$. For $J_H=0$ we obtain
negative values of $\Gamma$ (not shown in Fig.~\ref{fig:kondoxn}), with
$\Gamma \to -1/4$ in the limit $n \to 0$ [see Eq.~(\ref{eq:gamn})]. This is
consistent with previous results\cite{Alaor,Lacroix-2001} showing that
$\Gamma$ changes sign as a function of $n$ when the Kondo effect is strongly
dominant.

\begin{figure}
\includegraphics[width=8.5cm,clip]{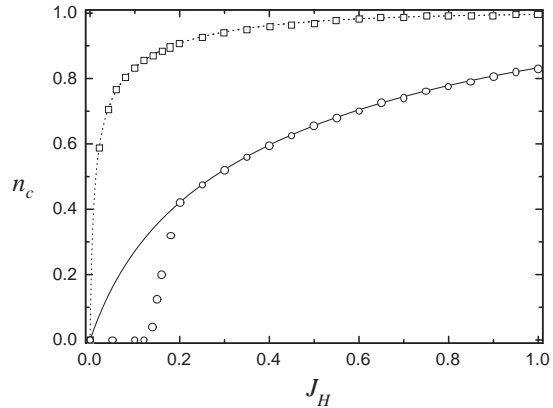}
\caption{ \label{fig:phdiag}Phase diagram plotted as the critical band
  filling $n_{c}$ versus $J_{H}$.  The curves are drawn from
Eq.~(\ref{eq:nc}) for $J_K/D = 0.4$ (dotted line) and 1.0 (solid line). The
symbols correspond to the results obtained by minimizing the energy. In each
case, the Kondo regime is stable above the line, and the magnetic phase
below.}
\end{figure}

Figure \ref{fig:phdiag} shows the phase diagram for $J_{K}/D=0.4$ and 1.0.
In the latter case (previously reported in Ref.~\onlinecite{Lacroix-2001}),
for small $J_{H}$ the Kondo phase is stable for all values of the band
filling, while for large $J_{H}$ the Kondo phase is stable only for $n >
n_{c}$ given by Eq.~(\ref{eq:nc}). Figure \ref{fig:phdiag} also shows that
Eq.~(\ref{eq:nc}) is very close to the numerical result in this
region, but does not apply when $J_H$ is very small, since the
assumed maximum in the lower energy band no longer exists. The crossover
between these two regimes was obtained numerically. In the crossover region
we found a {\em discontinous} transition from ${\lambda} {\neq} 0$ to
${\lambda} = 0$. These results have to be taken with caution, however, since
the value of $J_K/D =1$ is unphysically high, and the Kondo lattice is
expected to show ferromagnetic behavior in the low-$n$
limit,\cite{Tsunetsugu} while our analysis is restricted to
antiferromagnetic intersite exchange.

Thus, our calculation shows clearly that small $n$ and large $J_{H}$
values tend to suppress the Kondo effect, yielding a ``magnetic'' phase with
${\lambda}= 0$ and large short-range magnetic correlations.  In fact, in
this region a long-range magnetic order should certainly be stabilized, but
this was not taken into account in this approach. In contrast, both
${\lambda}$ and ${\Gamma}$ are different from zero in the Kondo phase.

\section{\label{sec:filling2}Band-filling effects at finite temperature}

Focusing on the Kondo-lattice problem at finite temperatures, the
half-filled case has been previously described, and we consider here the
general case of $n <1$. The number of $f$ electrons $n_{f}$ is always taken
equal to 1. The values of ${\lambda}$ and ${\Gamma}$ are determined by
self-consistently solving Eqs.~(\ref{eq:lamgam}) or by minimizing the free
energy, which is given by
\begin{eqnarray} \label{eq:F2}
F & = & -2\,T \sum_{k, \alpha=\pm }\ln \left[ 1+ e^{-(E_{\alpha}-\mu)/T}
\right]\\ \nonumber & & \mbox{} - E_{0} + zJ_{H} \Gamma ^{2} + 2J_{K}
\lambda^{2} \;.
\end{eqnarray}

In our mean-field approximation, $T_{K}$ and $T_{\rm cor}$ are defined as
the temperatures at which respectively ${\lambda}$ and $\Gamma$ become
zero. We have shown, in the case $n=1$, that the Kondo temperature $T_{K}$
for the lattice can be much smaller than the Kondo temperature $T_{K0}$ for
the single impurity, as observed in some cerium compounds.\cite{Iglesias}
Moreover, we have also shown that the temperature $T_{\rm cor}$ for the
occurrence of short-range magnetic correlations lies above the Kondo
temperature $T_{K}$ for large $J_{H}$ values, while $T_{\rm cor} = T_{K}$
for small $J_{H}$. These two theoretical results have provided a good
explanation for the previously described anomalous behavior observed in some
cerium compounds.

First, we present analytical results obtained in the case of large
$J_{H}$ values, where $T_{\rm cor} > T_{K}$. In this case, $T_{\rm
cor}$ is easily determined by taking the limit ${\Gamma} \to 0$ in
the self-consistent equation for ${\Gamma}$ with fixed ${\lambda}
= 0$, and we obtain
\begin{equation} \label{eq:Tcor}
T_{\rm cor}  =  zJ_{H}/12 \;,
\end{equation}
where $z$ is the number of nearest neighbors. In the simple cubic case, $ z
= 6 $, we obtain $T_{\rm cor} = J_{H}/2$.  Still for the case $T_{\rm cor} >
T_{K}$, we determine $T_{K}$ by taking the limit ${\lambda} \to 0$ in the
self-consistent equation for ${\lambda}$, with ${\Gamma}$ remaining
finite. We obtain the following equation to determine $T_{K}$:
\begin{eqnarray} \label{eq:oneove}
\frac{1}{\rho J_K} & = & \int_{-D}^{-D}
d\varepsilon \left[\frac{1}{\varepsilon(1-B)-E_0} \nonumber \times
\right. \\ & & \left.\left(
\frac{1}{1+e^{\frac{\varepsilon-\mu}{T_K}}}-
\frac{1}{1+e^{\frac{E_0 + B \varepsilon - \mu}{T_K}}}\right)\right] \;.
\end{eqnarray}
Equation (\ref{eq:oneove}), together with the self-consistent evaluation of
$B$ [see Eqs.~(\ref{eq:zJH}) and (\ref{eq:lamgam})], and the two equations
giving $n_f = 1$ and $n <1 $ (which yield $E_0$ and $\mu$) allow us to
determine $T_K$. In the case of $J_H=0$, we recover the simple equation
\begin{equation} \label{eq:twoove}
\frac{2}{\rho J_K} = \int_{-D}^{-D}d\varepsilon
\left[\frac{1}{\varepsilon - E_0} \tanh{\left(\frac{\varepsilon - \mu}{2
T_K}\right)}\right] \;,
\end{equation}
which reproduces the single-impurity result. In the limit $J_K/D << 1$,
this yields
\begin{equation} \label{eq:simple}
T_K = C_0 D \sqrt{n(2-n)} \exp{\left(-\frac{1}{\rho
J_K}\right)} \;,
\end{equation}
where $C_0$ is a numerical constant: $C_0 = 1.1337$.  Eq.~(\ref{eq:simple})
gives $T_K$ depending on the band filling as $\sqrt{n(2-n)}$, which is
consistent with a $\sqrt{n}$ dependence when $n \to 0$, as found in
Ref.~\onlinecite{Burdin}.

\begin{figure}
\includegraphics[width=8.5cm,clip]{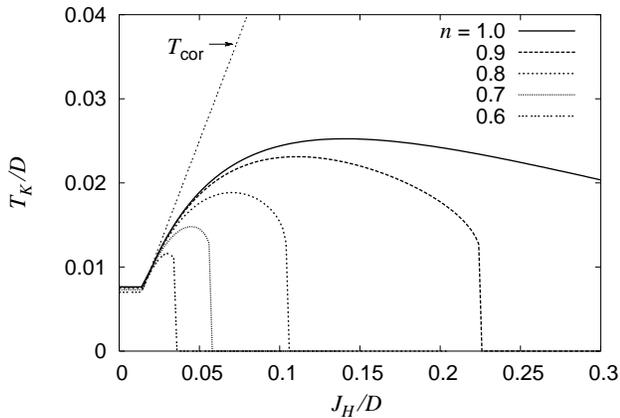}
\caption{\label{fig:TK}Kondo temperature $T_{K}$  versus $J_{H}$
  for $J_{K}/D=0.4$ and several values of $n$. We also show the correlation
temperature $T_{\rm cor}$.}
\end{figure}

\begin{figure}
\includegraphics[width=8.5cm,clip]{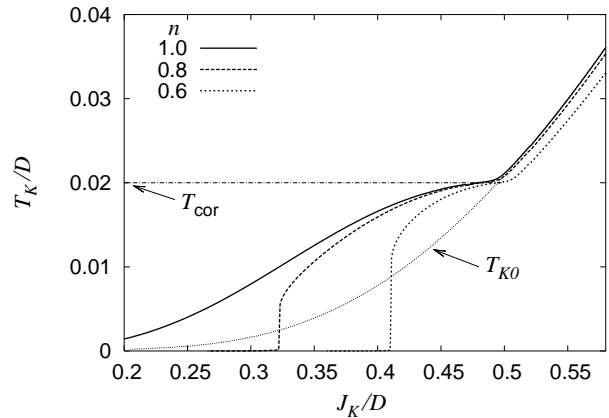}
\caption{\label{fig:TKTcor}Kondo temperature $T_{K}$ as a
  function of $J_{K}$ for $J_{H}/D=0.04$ and representative values of
  $n$. We also show the correlation temperature $T_{\rm cor}$, and the
  single-impurity Kondo temperature $T_{\rm K0}$.}
\end{figure}

Now, we present numerical results obtained at finite temperatures in the
case $n<1$, for different sets of parameters $J_{K}$, $J_{H}$ and $n$, in
order to study the Kondo-lattice properties and band-filling effects. Figure
\ref{fig:TK} gives the curves of the Kondo temperature $T_{K}$ versus
$J_{H}$ for a given $J_{K}$ value and several values of the conduction-band
filling $n$. We see clearly that $T_{K}$ first increases, and then decreases
with $J_{H}$ for fixed $n$, dropping abruptly to zero at some critical value
of $J_H$. On the other hand, for a given $J_H$, $T_K$ decreases rapidly as
$n$ departs from half filling. The enhancement of $T_K$ for small $J_H$ was
masked in Ref.~\onlinecite{Iglesias} by the choice of $J_H = \alpha J_K^2$,
which intended to mimic the RKKY interaction. Besides this being a
questionable approximation, we have preferred here to study the behavior of
the Kondo temperature as a function of $J_K$ and $J_H$ considered as
independent parameters. For comparison, in Fig.~\ref{fig:TK} we have also
plotted the correlation temperature $T_{\rm cor}$, which is linear with
$J_{H}$ independently of the value of $n$.\cite{Iglesias} Figure
\ref{fig:TKTcor} gives $T_{K}$ as a function of $J_{K}$ for $J_{H}/D=0.04$
and representative values of $n$. Here again we include the correlation
temperature $T_{\rm cor}$, which signals the onset of short-range magnetic
correlations when the temperature is lowered at fixed $J_K$. For comparison,
we plot the single-impurity Kondo temperature $T_{K0}$ (for $n \lesssim 1$),
which varies exponentially with $J_K$, and is weakly dependent on $n$ near
half-filling [see Eq.~(\ref{eq:simple})].

Figures \ref{fig:TK} and \ref{fig:TKTcor} show some interesting results of
our model. First, one can see the occurrence of short-range magnetic
correlations above the Kondo temperature, in good agreement with
experiment. Also, in the region of coexistence between Kondo effect and
magnetic correlations, the Kondo temperature is enhanced with respect
to the single-impurity case, and shows a smoother variation with $J_K$. Such
an enhancement agrees with the idea\cite{Coleman} that $f$-$f$ correlations
may, in some cases, reinforce the Kondo effect on a lattice. The
second noticeable feature of Figs.~\ref{fig:TK} and \ref{fig:TKTcor} is the
almost catastrophic supression of the Kondo effect with increasing intersite
coupling, and the enhancement of this behavior as the band-filling factor is
reduced.

Now, in order to better understand the physics of the Kondo lattice at low
temperatures, and to see more clearly the different regimes occurring with
decreasing temperature, we present different curves showing the product of
magnetic susceptibility and temperature, $\chi\,T$, which represents the
square of the effective magnetic moment.

\begin{figure}
\includegraphics[width=8.5cm,clip]{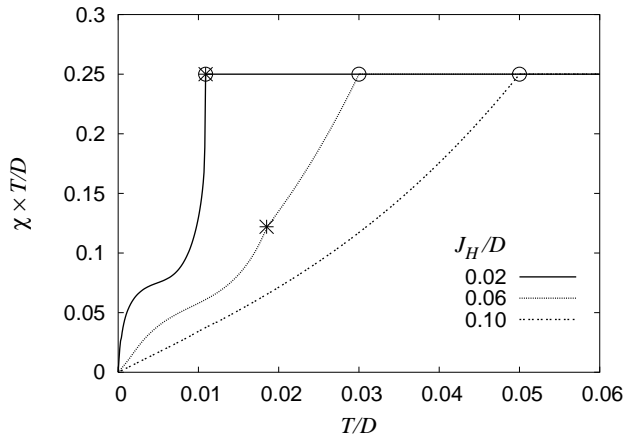}
\caption{Magnetic susceptibility times temperature
as a function of temperature for $n=0.8$, $J_{K}/D=0.4$, and different
values of $J_{H}$. The open circles mark the position of $T_{\rm cor}$, and
the stars, that of $T_K$. These two temperatures coincide in the case of
$J_H/D = 0.02$.} \label{fig:SusxT1}
\end{figure}

\begin{figure}
\includegraphics[width=8.5cm,clip]{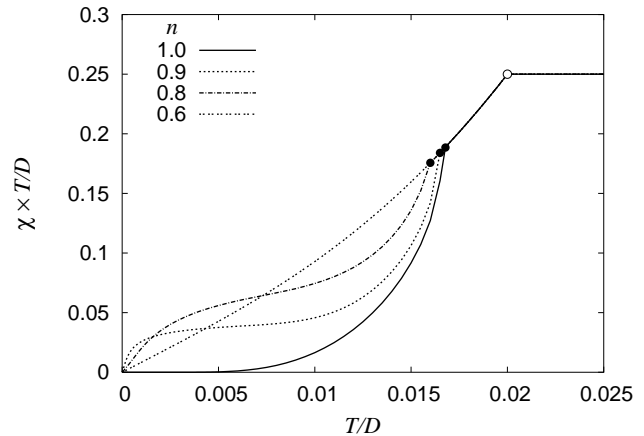}
\caption{Magnetic susceptibility times temperature as a function of
  temperature for $J_{K}/D=0.4$, $J_{H}/D=0.04$ and different values of
$n$. The open circle signals $T_{\rm cor}$, while the filled circles
indicate the positions of $T_K$ for the three higher values of $n$ (the
Kondo regime no longer occurs for $n=0.6$).} \label{fig:SusxT2}
\end{figure}

On the two following figures, $\chi\,T$ is constant at high temperatures
down to a first kink corresponding to either the correlation temperature
$T_{\rm cor}$ for a sufficiently large $J_{H}$ value or to $T_{K}$ at low
$J_{H}$ values. The kink corresponding to $T_{K}$ in the case when
$T_{K}<T_{\rm cor}$ is less pronounced, but still clearly visible. Figure
\ref{fig:SusxT1} gives the results for the temperature dependence of the
product $\chi T$ for $J_{K}/D=0.4$, $n=0.8$, and representative values of
$J_{\rm H}$.  This figure shows a Kondo regime alone for small $J_{H}$, a
regime with only short-range magnetic correlations above the Kondo regime
for intermediate $J_{H}$, and the absence of the Kondo phase for large
$J_H$. Then, Fig.~\ref{fig:SusxT2} gives the results for the temperature
dependence of the product $\chi T$ for $J_{K}/D=0.4$, an intermediate value
$J_{H}/D=0.04$, and representative values of $n$. The curves show the onset
of short-range correlations and the Kondo regime for $n$ values not very far
from 1, while the Kondo effect has completely disappeared for $n=0.6$, in
which case a single critical temperature remains. We immediately see on this
figure that the behavior of the magnetic susceptibility is completely
different for $n=1$ and smaller $n$ values, because the $n=1$ case
corresponds to an insulating system, and smaller $n$ values yield metallic
behavior. We also see that the Kondo effect disappears when $n$ decreases
below a critical value, which can be interpreted as a manifestation of the
``exhaustion'' problem\cite{Nozieres} in connection with the competition
between Kondo effect and magnetic correlations.

\begin{figure}
\includegraphics[width=8.5cm,clip]{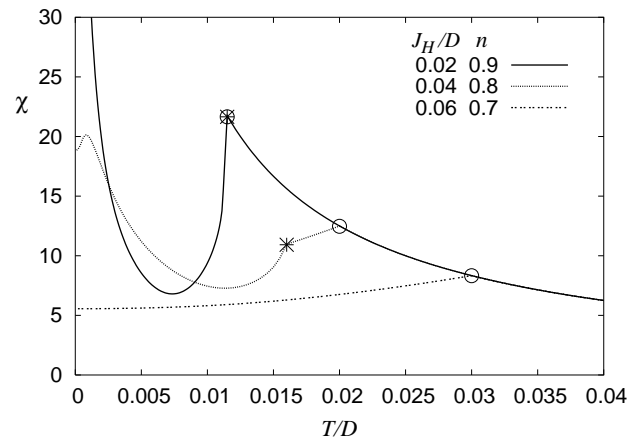}
\caption{Magnetic susceptibility as a function of temperature for
  $J_{K}/D=0.4$ and different sets of the parameters $J_{H}$ and $n$, in
order to show the different regimes. Circles and stars follow the convention
of Fig.~\ref{fig:SusxT1}.}
\label{fig:Sus}
\end{figure}

Finally, in figure \ref{fig:Sus}, we present a plot of the magnetic
susceptibility versus temperature for the three different regimes for
typical sets of parameters: Kondo regime alone (small $J_H$, $n$ near
half-filling), both magnetic correlations and Kondo regimes (higher $J_H$,
smaller $n$), and short-range magnetic correlations only (still higher
$J_H$, relatively small $n$).  This figure gives an illustrative summary of
the physics of the Kondo-lattice model.

The different behaviors shown in the last three figures are worth analyzing
in more detail. At high temperatures, the magnetic susceptibility is
decreasing in $1/T$ as normally expected. Then, below the highest
characteristic temperature ($T_{\rm cor}$ or $T_{K}$), there is a rapid
decrease of the product $\chi T$, indicating a reduction of the effective
magnetic moment due to the occurence of short-range antiferromagnetic
correlations, which tend to reduce the response of the localized moments to
an applied magnetic field, and an even faster reduction due to the Kondo
effect. At lower temperatures, within the Kondo regime, we observe a
flattening of the effective magnetic moment. This region of nearly constant
magnetic moment below $T_K$ can be related to the ``exhaustion'' problem,
because there are not enough electrons to screen the localized spins. At
still lower temperatures, a coherent regime sets in, and this residual
moment becomes completely screened. This behavior is not observed in the
half-filled case, where the effective magnetic moment tends very rapidly to
zero due to the existence of a gap at the Fermi level.

\section{\label{sec:concl}CONCLUSIONS}

We have studied the different behaviors of the intra-site and
inter-site correlations, and the characteristic temperatures as a
function of the intrasite Kondo exchange coupling, of the
intersite exchange coupling, and of the number of conduction
electrons relative to the number of $f$ electrons, which remains
fixed to 1. We use here a mean-field approximation with two order
parameters in each case, the mean-field Kondo correlator and the
short-range magnetic correlation.

We have established here for $n<1$, as in previous
work\cite{Iglesias} for $n=1$, that the dependence of the Kondo
temperature with the coupling constant $J_K$ for the lattice can
be significantly different from the single-impurity case. This
result can account for the pressure dependence of $T_{K}$ observed
in CeRh$_{2}$Si$_{2}$,\cite{Graf}
CeRu$_{2}$Ge$_{2}$,\cite{Sullow,Wilhelm} or more recently
Ce$_{2}$Rh$_{3}$Ge$_{5}$.\cite{Umeo} On the other hand, depending
on the relative values of $J_H$ and $J_K$, as well as on the
band-filling, the lattice Kondo temperature can closely follow the
single-impurity one, as observed in many cerium and all ytterbium
compounds. Further experiments are needed to better understand the
conditions yielding a Kondo temperature for the lattice much
different than the single-impurity one. This issue has also been
addressed by different theoretical approaches to both the Kondo
lattice and the Periodic Anderson
Model.\cite{Burdin,Castro,Ben-Hur,Continentino, Tahvildar, TSUrev}

Another interesting result concerns the derivation of a {\em
correlation temperature} below which short-range magnetic
correlations appear, in good agreement with neutron scattering
experiments in cerium compounds. These correlations can coexist
with the Kondo effect, but eventually dominate, and suppress the
Kondo regime for sufficiently high values of the intersite
exchange interaction or sufficiently low band fillings.

Our present calculation addresses again the difficult issue of the
nature of the ground state and screening in the Kondo-lattice
problem. We have shown here that, as the number of conduction
electrons is reduced, exhaustion may be compensated by formation
of intersite singlets of localized spins. Exact calculations for
small clusters would be interesting in that a comparison between
results with a number of conduction electrons equal to, or much
smaller than the number of 4$f$ electrons localized on the
different sites of the clusters could shed new light on this
issue.\cite{Busser} Finally, it is interesting to notice that
taking into account lattice effects is essential for describing
the properties of cerium or other anomalous rare-earth compounds
at low temperatures, as shown, for example, in photoemission
experiments.\cite{Tahvildar,Arko}


\begin{thebibliography}{99}
\bibitem{Doniach} S. Doniach, Proceedings of the ``Int. Conf. on Valence
Instabilities and Related Narrow-band Phenomena'', ed. By R. D.  Parks,
Plenum Press, 168 (1976).
\bibitem{Barbara} B. Barbara, H. Bartholin, D. Florence, M.F. Rossignol and
E.  Walker, Physica B {\bf 86-88}, 177 (1977).
\bibitem{Graf} T. Graf, J. D. Thompson, M. F. Hundley, R. Movshovich,
Z. Fisk, D.  Mandrus, R. A. Fischer and N. E. Phillips,
Phys. Rev. Lett. {\bf 78}, 3769 (1997).
\bibitem{Alami} K. Alami-Yadri, H. Wilhelm and D. Jaccard, Physica B {\bf
259-261}, 157 (1999).
\bibitem{Sullow} S. Sullow, M.C. Aronson, B.D. Rainford and P. Haen,
Phys. Rev.  Lett. {\bf 82}, 2963 (1999)
\bibitem{Wilhelm} H. Wilhelm, K. Alami-Yadri, B. Revaz and D. Jaccard,
Phys. Rev. B {\bf 59}, 3651 (1999).
\bibitem{Iglesias} J. R. Iglesias, C. Lacroix and B. Coqblin, Phys. Rev.{\bf
B 56}, 11820 (1997).
\bibitem{Umeo} K. Umeo, T. Takabatake, T. Suzuki, S. Hane, H. Mitamura and
T.  Goto, Phys. Rev. B {\bf 64}, 144412 (2001).
\bibitem{Rossat} J. Rossat-Mignod, L. P. Regnault, J. L. Jacoud, C. Vettier,
P.  Lejay, J. Flouquet, E. Walker, D. Jaccard and A. Amato,
J. Magn. Magn. Mater. {\bf 76-77}, 376 (1988).
\bibitem{Pierre} J. Pierre, P. Haen, C. Vettier and S. Pujol, Physica B {\bf
163}, 463 (1990).
\bibitem{Flouq} J. Flouquet, S. Kambe, L. P. Regnault, P. Haen, J.P. Brison,
F.  Lapierre and P. Lejay, Physica B {\bf 215}, 77 (1995).
\bibitem{Regnault} L. P. Regnault, W. A. C. Erkelens, J. Rossat-Mignod,
P. Lejay, J.  Flouquet, Phys. Rev. B {\bf 38}, 4481 (1988).
\bibitem{Mignot} J. M. Mignot, J. L. Jacoud, L. P. Regnault,
J. Rossat-Mignod, P.  Haen, P. Lejay, P. Boutrouille, B. Hennion and
D. Petitgrand, Physica B {\bf 163}, 611 (1990).
\bibitem{Regnault2} L. P. Regnault, J. L. Jacoud, J. M. Mignot,
J. Rossat-Mignod, C.  Vettier, P. Lejay and J. Flouquet, Physica B {\bf
163}, 606 (1990).
\bibitem{Lacroix-79} C. Lacroix and M. Cyrot, Phys. Rev. B {\bf 20}, 1969
(1979).
\bibitem{Nozieres} P. Nozi\`eres, Eur. Phys. J. B {\bf 6}, 447 (1998).
\bibitem{Alaor} A. R. Ruppenthal, J. R. Iglesias and M. A. Gusm\~{a}o, Phys. Rev
{\bf B 60} 7321, (1999).
\bibitem{Coq-2000} B. Coqblin, M. A. Gusm\~ao, J. R. Iglesias, C. Lacroix,
A.  Ruppenthal and Acirete S. Da R. Simoes, Physica B {\bf 281-282}, 50
(2000).
\bibitem{Coq-2001} B. Coqblin, M. A. Gusm\~ao, J. R. Iglesias, A. Ruppenthal
and C.  Lacroix, J. Magn. Magn. Mater.{\bf 226-230}, 115 (2001).
\bibitem{Jullien} R. Jullien and P. Pfeuty, J. Phys. F {\bf 11}, 353 (1981).
\bibitem{Continentino} M. A. Continentino, G. M. Japiassu and A. Troper
Phys. Rev. B {\bf 39}, 9734 (1989).
\bibitem{Tsunetsugu} H. Tsunetsugu, Y. Hatsugai, K. Ueda and M. Sigrist,
Phys. Rev. B {\bf 46}, 3175 (1992).
\bibitem{Burdin} S. Burdin, A. Georges and D. R. Grempel,
Phys. Rev. Lett. {\bf 85}, 1048 (2000)
\bibitem{Burdin2} S. Burdin, Thesis, Grenoble (2001).
\bibitem{Castro} A.H. Castro Neto and B.A. Jones, Phys. Rev. B {\bf 62},
14975 (2000).
\bibitem{Ben-Hur} B. H. Bernhard, C. Lacroix, J. R. Iglesias and B. Coqblin,
Phys. Rev.  B {\bf 61}, 441 (2000).
\bibitem{Tahvildar} A. N. Tahvildar, M. Jarrell and J. K. Freericks,
Phys. Rev.  B {\bf 55}, R3332 (1997).
\bibitem{Arko} A.J. Arko, J.J. Joyce, A.B. Andrews, J.D. Thompson,
J.L. Smith, D. Mandrus, M.F. Hundley, A.L. Cornelius, E. Moshopoulou,
Z. Fisk, P.C. Canfield and Alois Menovsky, Phys. Rev.B {\bf 56}, R7041
(1997).
\bibitem{Lacroix-2001} C. Lacroix, J. R. Iglesias, and B. Coqblin, Physica B
{\bf 312-313}, 159 (2002).
\bibitem{Coleman} P. Coleman and N. Andrei, J. Phys. Condens.  Matter, {\bf
1}, 4057 (1989).
\bibitem{TSUrev} H. Tsunetsugu, M. Sigrist, and K. Ueda,
 Rev. Mod. Phys. {\bf 69}, 809 (1997).
\bibitem{Busser} C. Busser, E. V. Anda, J. R. Iglesias, and B. Coqblin,
J. Magn. Magn. Mater. {\bf 226-230}, 134 (2001).
\end{thebibliography}
\end{document}